# Oblivious Transfer using Elliptic Curves

Abhishek Parakh
Louisiana State University, Baton Rouge, LA

May 24, 2006

***Abstract:*** *This paper proposes an algorithm for oblivious transfer using elliptic curves. Also, we present its application to chosen one-out-of-two oblivious transfer.*

## 1  Introduction

An oblivious transfer scheme is a protocol in which a sender sends a message to a receiver with some fixed probability between 0 and 1 without the sender knowing whether or not the receiver received the message. The idea was introduced in 1981 by Michael Rabin [1], [2].

Rabin developed a solution to the problem of mutual exchange of secrets between two distrustful parties. For example, Alice and Bob have secrets $S_A$ and $S_B$, respectively, which they wish to exchange ($S_A$ may be the password to a file that Bob wants and vice versa). The problem is to establish a protocol without a trusted third party and without simultaneous exchange of messages.

Rabin exploited the fact that a square transformation $c = m^2 \mod n$, $n = p \times q$, $p$ and $q$ are primes, results in two or four messages being mapped to a single cipher. Hence, using Rabin's protocol, Alice would convey the factors of $n_A$ (assuming Alice is using a public key encryption method of the form $c = m^e \mod n_A$, where $e$ is the encryption exponent) without knowing for sure whether Bob received the factors or not. In other words, Bob may or may not receive the factors, each happening with probability one-half.

In this paper we introduce an oblivious transfer protocol using elliptic curve cryptography (ECC), a fast upcoming competitor against RSA. We use exactly the same set up as described in [2]. We present an algorithm that achieves oblivious transfer between two parties using elliptic curves for encryption of their messages. Section 2 discuses the basics of elliptic curve cryptography. Section 3 introduces the key observation that led to the idea of oblivious transfer and Section 4 presents our algorithm together with an illustrative example. Section 5 presents an application of our oblivious transfer algorithm to chosen one-out-of-two oblivious transfer.

## 2 Basics of elliptic curves

An elliptic curve used for cryptographic purposes is defined as follows:

$$y^2 = x^3 + ax + b \tag{1}$$

where $a$ and $b$ are integer constants. The set of points $E(a,b)$ is a set $(x, y)$ of all $x$ and $y$ satisfying (1).

For an elliptic curve over a finite field $Z_p$, we use the cubic equation (1) in which the variables and coefficients all take on values in the set of integers from 0 and $p-1$, for some prime $p$, in which calculations are performed modulo $p$. Thus, we use

$$y^2 \bmod p = (x^3 + ax + b) \bmod p \tag{2}$$

for cryptographic applications over finite fields. This set of points is denoted as $E_p(a,b)$. The order $n$ of a point $G = (x_1, y_1)$ on an elliptic curve is defined as the smallest positive integer $n$ such that $nG = 0$. Point $G$ is called the base point in $E_p(a,b)$ and is picked such that its order $n$ is a very large value.

The security of ECC arises from the fact that for $Q = kP$, where $Q, P \in E_p(a,b)$ and $k < p$, it is easy to calculate $Q$ given the values of $k$ and $P$, but it is relatively very hard to determine $k$ given the values of $Q$ and $P$.

A standard elliptic curve transfer proceeds as follows: the first task in this system is to encode the plain text message $m$ to be sent as a $x-y$ point $P_m$. It is the point $P_m$ that will be encrypted as cipher-text and subsequently decrypted. We cannot simply encode the message as the $x$ or $y$ coordinate at a point, because not all such coordinates are in $E_p(a,b)$. Each user A selects a private key $n_A$ and generates a public key $P_A = n_A \times G$. To encrypt and send a message $P_m$ to B, A chooses a random positive integer $k$ and produces a cipher-text $C_m$ consisting of the pair of points

$$C_m = \{ kG ; \ P_m + kP_B \}$$

A has used B's public key $P_B$. To decrypt the cipher-text, B multiplies the first point in the pair by B's secret key and subtracts the result from the second point:

$$P_m + kP_B - n_B(kG) = Pm + k(n_B G) - n_B(kG) = P_m$$

Note that A has masked the message $P_m$ by adding $kP_B$ to it. Nobody but A knows the value of $k$, so even though $P_B$ is public, nobody can remove the mask $kP_B$. Reader may refer to [3] for further background on elliptic curve cryptography.

## 3    Key Observation

If we look closely at square transformation in [1] and the elliptic curve equation given by (2), we can rewrite (2) as

$$y^2 \bmod p = S \tag{3}$$

where $S = (x^3 + ax + b) \bmod p$. It should be clear to the reader that for every $x$ coordinate there are two possible $y$ coordinates. However, unlike in square transformation, here neither $x$ nor $y$ can be substituted for message, because not all values of $x$ and $y$ are permissible in ECC.

## 4    The Proposed Algorithm

Our aim is to allow exchange of secret $S_A$ and $S_B$ between two parties A and B without using a trusted third party and without simultaneous exchange. Here, we do not go into the details of signing the messages using ECC and take it for granted that all the messages are signed.

Both A and B select a common elliptic curve $E_q(a,b)$. This information is public. They then decide upon one $x$-coordinate. Let the two points corresponding to this $x$-coordinate be $P_1$ and $P_2$, whereupon by symmetry $P_1 = -P_2$. The $x$-coordinate is also public knowledge. Since, A and B have not decided upon which $y$-coordinate to use, we will denote A's choice of point as $P_A$ and B's choice as $P_B$, such that

$$P_A = P_1 \quad \text{or} \quad P_A = P_2$$

Similarly, $\quad P_B = P_1 \quad$ or $\quad P_B = P_2$.

Even though the $x$-coordinate is common, neither party knows what is the final point chosen by the other because there are two possible $y$-coordinates to choose from.

Now, let A choose a secret key $n_A$, which she wishes to use for encryption of her messages, with the aim of obliviously conveying this secret key $n_A$ to B. Also, we assume that a procedure for mapping of $n_A$ to a point on elliptic curve has been pre-decided. We call the point on our elliptic curve, corresponding to $n_A$, as $P_{n_A}$. Thus, if a person knows $P_{n_A}$, he can deduce $n_A$ from it. Similar, arrangement is made on B's side too.

Under the above assumptions, the oblivious transfer of secret key proceeds as follows:

1. A sends to B : $n_A P_A$

2. B sends to A : $\{ n_B P_B ; n_B(n_A P_A) + R ; n_B R \}$

   where, $n_B$ is B's secret key
   
   $R$ is randomly chosen point by B, belonging to the group $E_q(a,b)$.

3. A does : $n_A [ n_B(n_A P_A) + R - n_A(n_B P_B) ] = Q$

4. A sends to B : $\{ n_A(n_B P_B) + Q ; n_A(n_B R) + P_{n_A} \}$

5. B does :
   a. $n_A(n_B P_B) + Q - n_B(n_A P_A) = K$
   
   b. $n_A(n_B R) + P_{n_A} - n_B K = Z_B$

The sequence of steps presented above achieves our goal of oblivious transfer. The two cases that arise in such a transfer are $P_A = P_B$ and $P_A \neq P_B$. We discuss these two cases below and show how the algorithm given above achieves our goal.

The difference between the two cases arises from step 3. Hence, we analyze them step 3 onwards.

---

***Case I:*** $P_A = P_B$

3. A does : $n_A [ n_B(n_A P_A) + R - n_A(n_B P_B) ] = n_A R$

4. A send to B : $\{ n_A(n_B P_B) + n_A R ; n_A(n_B R) + P_{n_A} \}$

5. B does :
   a. $n_A(n_B P_B) + n_A R - n_B(n_A P_A) = n_A R$
   
   b. $n_A(n_B R) + P_{n_A} - n_B(n_A R) = P_{n_A}$

---

***Case II:*** $P_A \neq P_B$

In this case, we note that $P_A = -P_B$. Therefore, the results are as follows:

3. A does : $n_A [ n_B(n_A P_A) + R - n_A(n_B P_B) ] = n_A [ 2 \times n_B(n_A P_A) + R ]$

4. A send to B : $\{ n_A(n_B P_B) + n_A[ 2 \times n_B(n_A P_A) + R ]; n_A(n_B R) + P_{n_A} \}$

5. B does :
   a. $n_A(n_B P_B) + n_A[ 2 \times n_B(n_A P_A) + R ] - n_B(n_A P_A) = K$
   
   b. $n_A(n_B R) + P_{n_A} - n_B(K) \neq P_{n_A}$

Once the receiver knows $P_{n_A}$, he can deduce $n_A$ from it. Therefore, this point forward we refer to $P_{n_A}$ as $n_A$.

However, it is to be noted that no matter what calculations are performed by B in step 5, he cannot get $n_A$ if $P_A \neq P_B$. The problem is equivalent to the discrete log problem in case of $P_A \neq P_B$.

Since, $P_A = P_B$ with probability one-half, B receives the secret key $n_A$ with probability one-half.

Returning to our algorithm, B can verify the value $Z_B$ it has obtained from step 5, whether it is $n_A$ or not, by doing $Z_B \times P_1$ and $Z_B \times P_2$ and checking if one of them is equal to $n_A P_A$ sent to it by A in the first step.

In a similar manner, B transfers its secret key $n_B$ to A with probability one-half. Once this transfer has been achieved we can follow similar step proposed in [2] in order to prevent cheating by either of the parties during exchange of information. Here, we present these steps, adapting them to suit elliptic curve transfers. We define the state of knowledge of the secret keys as follows:

$$k_A = \begin{cases} M, & \textit{if A knows B's secret key} \\ \overline{M}, & \textit{if A does not know B's secret key} \end{cases}$$

Similarly,

$$k_B = \begin{cases} M, & \textit{if B knows A's secret key} \\ \overline{M}, & \textit{if B does not know A's secret key} \end{cases}$$

where $M$ is a constant and $\overline{M}$ is the bit wise complement of $M$.

After the transfer of keys according the algorithm presented in this paper and having defined the state of knowledge of keys as above,

A sends to B: $k_A \oplus S_A$

B sends to A: $k_B \oplus S_B$

Note that the above two steps do not provide either party any information about other's secret. Now, A may transfer its secret to B using an elliptic curve cryptographic transfer. However, A will encode the secret using its own secret key and not the public key of the other party, as is usually done in a standard elliptic curve transfers; G is the base point with large order.

A sends to B : $S_A + n_A G$

B does (assuming he knows $n_A$) : $S_A + n_A G - n_A G = S_A$

B transfers its secret to A in the next step in a similar manner. However, suppose, at the last step B were to cheat and not pass on his secret $S_B$ to A, then the fact that B has cheated A implies that B has $n_A$, i.e. $k_B \oplus S_B = M \oplus S_B$.

Thus, A can do $M \oplus S_B \oplus M = S_B$ and thus obtain $S_B$. The probability, when the protocol is completed, that neither one knows other's secret is one-quarter.

**Example:** Let A and B choose an elliptic curve $E_{23}(9, 21)$. The equation corresponding to this curve is $y^2 \mod 23 = (x^3 + 9x + 21) \mod 23$. Now, both parties decide upon a common $x$-coordinate, say 7. The two points corresponding to this $x$-coordinate are $P_1 = (7, 6)$ and $P_2 = (7, 17)$. From properties of elliptic curve, we have $P_1 = -P_2$.

Let A choose a secret number $n_A = 5$. We do not explore the details of its mapping of $n_A$ to the elliptic curve and just refer to it as $P_{n_A}$. In turn, let B chooses a secret number $n_B = 3$ and a random point $R = (2, 1)$. Now we execute our algorithm by considering the two cases separately:

***Case1:*** $P_A = (7, 6)$ and $P_B = (7, 6)$

1. A sends to B: $n_A P_A = 5(7, 6) = (11, 18)$.

2. B sends to A:
   $\{n_B P_B; \; n_B(n_A P_A) + R; \; n_B R\} = \{3(7, 6); \; 3(11, 18) + (2, 1); \; 3(2, 1)\}$
   $= \{(1, 10); \; (11, 5); \; (14, 19)\}$

3. A does:
$$n_A[\,n_B(n_A P_A) + R - n_A(n_B P_B)\,] = Q$$
$$= 5[\,(11, 5) - 5\,(1, 10)\,]$$
$$= 5[\,(11, 5) - (13, 9)\,]$$
$$= 5[\,(2, 1)\,] = (7, 17)$$

4. A sends to B:
$$\{\,n_A(n_B P_B) + Q;\ \ n_A(n_B R) + P_{n_A}\,\} = \{\,(13, 9) + (7, 17);\ \ 5\,(14, 19) + P_{n_A}\,\}$$
$$= \{\,(15, 9);\ (1, 13) + P_{n_A}\,\}$$

5. B does:
   a) $n_A(n_B P_B) + Q - n_B(n_A P_A) = K$
$$= (15, 9) - 3(11, 18)$$
$$= (15, 9) - (13, 9)$$
$$= (7, 17)$$

   b) $n_A(n_B R) + P_{n_A} - n_B(K) = (1, 13) + P_{n_A} - 3\,(7, 17)$
$$= (1, 13) + P_{n_A} - (1, 13)$$
$$= P_{n_A}$$

*Case 2:* $P_A = (7, 6)$ and $P_B = (7, 17)$

1. A sends to B: $n_A P_A = 5\,(7, 6) = (11, 18)$.

2. B sends to A:
$$\{\,n_B P_B;\ \ n_B(n_A P_A) + R;\ \ n_B R\,\} = \{\,3\,(7, 17);\ \ 3\,(11, 18) + (2, 1);\ \ 3\,(2, 1)\,\}$$
$$= \{\,(1, 13);\ (11, 5);\ (14, 19)\,\}$$

3. A does:
$$n_A[\,n_B(n_A P_A) + R - n_A(n_B P_B)\,] = Q$$
$$= 5[\,(11, 5) - 5\,(1, 13)\,]$$
$$= 5[\,(11, 5) - (13, 14)\,]$$
$$= 5[\,(3, 11)\,] = (9, 7)$$

4. A sends to B:
$$\{\,n_A(n_B P_B) + Q;\ \ n_A(n_B R) + P_{n_A}\,\} = \{\,(13, 14) + (9, 7);\ \ 5\,(14, 19) + P_{n_A}\,\}$$
$$= \{\,(17, 2);\ (1, 13) + P_{n_A}\,\}$$

5. B does:

a) $n_A(n_B P_B) + Q - n_B(n_A P_A) = K$
$$= (17, 2) - 3(11,18)$$
$$= (17, 2) - (13, 9)$$
$$= (2, 22)$$

b) $n_A(n_B R) + P_{n_A} - n_B(K) = (1, 13) + P_{n_A} - 3(2, 22)$
$$= (1, 13) + P_{n_A} - (14, 4)$$
$$\neq P_{n_A}$$

The above example makes the working of our algorithm clear.

## 5 Chosen one-out-of-two oblivious transfer

The chosen one-out-of-two oblivious transfer, $\binom{2}{1}$- OT for short, is an important application of the basic oblivious transfer protocol. In this transfer, the sender sends two secrets $s_0$ and $s_1$ and the receiver's input is choice bit $c$; the latter then learns $s_c$ but gets no information about other secret $s_{1-c}$.

This transfer has been implemented using exponentiations. Here we show that the one-out-of-two oblivious transfer can be implemented using the algorithm we presented.

We assume that both parties are willing to take part in the protocol honestly, i.e. A is willing to disclose one out of two secrets that it has to B, but B does not want A to know which one secret it wants to know. Also, B should learn only the one secret it wants to know and nothing about the other.

A is said to have two secrets $s_0$ and $s_1$. A, associates two different secret keys with each of them. These secret keys will be used to encrypt $s_0$ and $s_1$ when transferring them to B. B must be able to retrieve only one of these two secrets and A should not come to know, what B has extracted.

Let A associate keys $n_{A0}$ with $s_0$ and $n_{A1}$ with $s_1$ for encryption. Now, B's task is to retrieve one of these two keys, i.e. retrieve $n_{A0}$ if it wants to know $s_0$ and retrieve $n_{A1}$ if it wants to know $s_1$, in such a manner that A should not come to know which key B retrieved and B should not gain any information about the other key associated with the other secret.

Recall, from the previous section that every $x$-coordinate yields two points $P_1$ and $P_2$ such that $P_1 = -P_2$. A declares that it is associating secret $s_0$ with point $P_1$ and secret $s_1$ with point $P_2$. The transfer of secret then proceeds as follows:

1. A sends to B : $\{ n_{A0}P_1 \ ; \ n_{A1}P_2 \}$

2. B sends to A : $\{ n_B P_B \ ; \ n_B(n_{A0}P_1) + R \ ; \ n_B(n_{A1}P_2) + R \ ; \ n_B R \}$

3. A does :
$$n_{A0} [ \ n_B(n_{A0}P_1) + R - n_{A0}(n_B P_B) \ ] = H_1 \ ;$$

$$n_{A1} [ \ n_B(n_{A1}P_2) + R - n_{A1}(n_B P_B) \ ] = H_2.$$

4. A sends to B :
$$\{ \ n_{A0}(n_B P_B) + H_1 \ ; \ n_{A0}(n_B R) + P_{n_{A0}} \ ; \quad n_{A1}(n_B P_B) + H_2 \ ; \ n_{A1}(n_B R) + P_{n_{A1}} \ \}$$

*Note*: $P_{n_{A0}}$ and $P_{n_{A1}}$ is the mapping of secret keys $n_{A0}$ and $n_{A1}$ to points on the elliptic curve.

B must have chosen $P_B$ in the second step such that $P_B = P_1$ if B wants secret $s_0$ and $P_B = P_2$ if B wants secret $s_1$. Therefore after step 4, B picks up only one of the two pairs of points sent to it by A which will yield it the secret key it wants.

For example, if B has chosen $P_B = P_1$ then the first pair of points in step 4, i.e. $\{ \ n_{A0}(n_B P_B) + H_1 \ ; \ n_{A0}(n_B R) + P_{n_{A0}} \ \}$, will yield $n_{A0}$ in the following manner :

5. B does :
   a) $n_{A0}(n_B P_B) + H_1 - n_B(n_{A0}P_1) = H_1 = n_{A0}R$
   b) $n_{A0}(n_B R) + P_{n_{A0}} + n_B(n_{A0}R) = P_{n_{A0}}$.

From $P_{n_{A0}}$, B can easily calculate $n_{A0}$. The second pair of points will not yield any key. Thus, B can get only one of the two secret keys and A remains oblivious to the fact that which of the two keys did B retrieve.

Now, A may send both the secrets to B in the following manner:

A sends to B: $\{ \ Ps_0 + n_{A0}G \ ; \ Ps_1 + n_{A1}G \ \}$, where $Ps_0$ is the mapping of secret $s_0$ to the elliptic curve.

B will be able to retrieve only $Ps_0$ in our example because it has only $n_{A0}$ and hence obtain $s_0$. It will not be able to get any information from the second half of the message about secret $s_1$. A does not know which of the two secrets did B obtain. We have achieved our goal of chosen one-out-of-two oblivious transfers.

## 6 Conclusions

In this paper we have introduced the idea of oblivious transfer to elliptic curves and presented an algorithm for its implementation. Also, we have shown how it can be applied to the traditional problem of $\binom{2}{1}$- OT.

The algorithm presented here may be expressed in different variants. The key contribution is the introduction of oblivious transfer to ECC. The one-out-of-two oblivious transfer may be further modified in order to obtain 1-out-of-n oblivious transfer.